\documentstyle[prl,aps,psfig]{revtex}
\begin{document}
\draft

\twocolumn[
\hsize\textwidth\columnwidth\hsize\csname@twocolumnfalse\endcsname\title
{Conductance of metallic carbon nanotubes dipped into a metal}

\author{N.Mingo$^1$ and Jie Han$^2$}
\address{ $^1$Mail Stop T27A-1\\$^2$Eloret Corp.\\NASA-Ames Research Center\\Moffett Field, CA 94035-1000}

\maketitle

\begin{abstract}                % DON'T CHANGE THIS LINE
We calculate the conductance variation of several metallic carbon nanotubes as their end is being dipped
into a liquid metal electrode, where experiments have shown an achievable conductance close to $2e^2/h$.
The calculated conductance for a (40,40) nanotube indicates that the current flows almost entirely 
through the $\pi$ mode, and not the $\pi^*$ mode.
The calculation also predicts that for narrower nanotubes ($\sim1nm$ in diameter),
in a 'weak coupling' regime, the saturation of both $\pi$ and $\pi^*$ modes
should be observable. An experiment is proposed to verify this point.
\\
\\
\bf{Accepted for publication in Phys.Rev.B rapid communications\\
\\}
\end{abstract}

]

\narrowtext
\begin{text}
A recent experiment \cite{Frank} measured the conductance of carbon nanotubes \cite{Saito} dipped into liquid 
metals, reporting long plateus of conductance close to $2e^2/h=1G_0$ (1 quantum of conductance) in a staircase shape.
These plateus are due to the dipping of consecutive nanotubes into the
liquid metal. Conduction is generally accepted to take place through the outermost wall in this case. 
In principle, two modes are available for this conduction to occur, namely the $\pi$ and $\pi^*$ modes, and
a conductance of $2G_0$ might be available. The fact that only $1G_0$ was reported was puzzling.
A subsequent calculation \cite{Ihm} for a (10,10) nanotube into a jellium, with a close jellium-carbon distance of
$d=0.7\AA$, suggested that conduction might be taking place via the $\pi^*$ mode, the $\pi$ mode being
totally reflected. However, neither the conductance dependence with dipped length, nor
larger diameters or different carbon-metal separations were investigated in ref. \cite{Ihm}.
A later study by Anantram \cite{Anantram}, performed in a flat geometry, 
argued that the conductance might be taking place via the $\pi$
mode instead. However, the conductances for each of the two modes were not shown separately there, and the geometry
was different from that in ref. \cite{Ihm}.

Here, for the first time, we address the problem of how the conductance varies with dipped length
as the tube is inserted into the metal, and how it saturates to an asymptotic value. 
Our study is for nanotubes up to 5.5 nm in diameter i.e. (40,40)
(the experiment used diameters $\gtrsim5nm$). We also consider the influence of the carbon-metal
separation. Despite the (10,10) case basically yields the same result as ref. \cite{Ihm} for $d=0.7\AA$, 
our analysis shows that the narrow (10,10) nanotube with a short carbon-metal separation 
is not an adequate description of the experimental case. Rather, a conduction via the $\pi$ mode (and not
the $\pi^*$) is obtained for the wider diameter tubes. The results are compatible with 
experimental conductance curves, and in disagreement with previous theoretical suggestion. Nevertheless,
the symmetry of the scattered electrons' wave functions in the metal region is a rapidly oscillating one in
all cases, in full agreement with the result in ref. \cite{Ihm}.
An experimental verification will be proposed at the end of the paper.

In order to study the conductance dependence with dipped length, $l$, we use two different configurations
that allow a better insight into the physical process. The "semi-infinite tube configuration", shown in fig.1a, corresponds
to a semi-infinite tube that is perpendicularly dipped a distance $l$ into a semi-infinite metal. 
The "infinite tube configuration", fig.1b,
corresponds to an infinite tube, traversing a slab of metal of width $l$. The slab is infinite in the
$x$ and $y$ directions. The experimental 
situation is most similar to the semi-infinite configuration. However, the infinite configuration is 
useful because we can calculate not just the total conductance, but also how much current goes into the
metal and how much travels through to the other side.

The metal electrons are regarded as free, with a Fermi energy of 7.13 eV corresponding to the free electron 
density of Hg. The free-electron versus Bloch-wave character of mercury has been discussed in ref. \cite{Hg}, and we will
not enter that issue here.
A cylindrical boundary to the free electrons
is set at a separation $d$ from the nanotube's carbon atoms (fig.1), 
and cylindrical coordinates are used to take full advantage of the symmetry.
This allows consideration of a single nanotube inserted in the metal,
and avoids spurious periodicities in the calculation.
The electrons at the nanotube are treated in an atomic orbital basis, and 
the tunneling matrix elements (see ref. \cite{Goldberg} for instance) from the carbon atoms 
to the electronic states in the metal are calculated for each case. 
The nanotubes are undoped, with the Fermi level determined by the charge neutrality
condition for the isolated nanotube. 

Results of conductance versus dipped length for
two different nanotubes will be shown: a (40,40) corresponding to the diameter used in the experiments, 
and a narrower (10,10). The Green function technique allows us to treat the semi-infinite metal 
and the infinite and semi-infinite nanotubes straightforwardly.
Using the Green functions, the conductances are also readily calculated \cite{Makoshi,Mingo}. Calculation of 
the partial conductances for the $\pi$ and $\pi^*$ modes is straightforwardly performed, since the isolated 
nanotube's Green functions for the two modes are completely independent (for non-capped nanotubes). A basis
transformation is applied to the atomic ring of the nanotube located just near the metal surface on its outer
side, yielding the Green function for the independent $\pi$ and $\pi^*$ modes of the non-submerged 
part of the nanotube, $g_{\pi,\pi}$ and $g_{\pi^*,\pi^*}$, with $g_{\pi,\pi^*}=0$. 
The Green function elements for the other modes are purely real at the Fermi level and do not contribute
to the current.
This Green function is then coupled to the submerged part of the nanotube, and the conductance from the two 
independent electron sources ($\pi$ and $\pi^*$) into the metal can be calculated by eqs.11-13 of ref. \cite{Mingo}.
The measurable conductance is just the sum of the $\pi$ and $\pi^*$ conductances.

All magnitudes in the problem are well known quantities, except for $d$, defined as the distance 
between the carbon nanotube and the jellium edge
(in ref.\cite{Ihm} $d$ was chosen to be $0.7\AA$, which is the covalent radius of the
carbon atom). Therefore we have performed calculations for different values of $d$. 
Comparison with experimental results might allow an estimation of the best suited value of $d$.
Despite the physical meaning of $d$ is clear, it is difficult to establish a direct 
relation between its numerical value and an average C-Hg atom distance, if a full atomic description
of the system were made instead of using a jellium. For the sake of comparison, if such non-jellium
calculation would be made in the future, the imaginary part of the self energy  at $E_F$ due to the metal,
projected onto a carbon atom's $p$ orbital perpendicular to it, is $\sim 0.5 eV$ for $d=1.5\AA$.

First of all, let us study nanotubes in the diameter range used in the experiments, $\sim 5 nm$.
For a 5.5 nm diameter armchair nanotube, i.e.(40,40), and $d=1.5\AA$, we plot the conductance of the two modes as
a function of the penetration length $l$, in fig.2-left. The $\pi$ mode conductance increases in an 
oscillatory fashion, saturating to near $0.95G_0$ when the nanotube is dipped deep enough into the metal.
The amplitude of the superimposed oscillations vanishes with increasing dipped length. On the other
hand, the conductance of the $\pi^*$ mode is smaller than $0.1G_0$ in the range from 0 to $120\AA$ dipped
length. The calculated conductance for the smaller radius nanotube -(10,10)- is shown in the same figure
(fig.2-right). Now, the $\pi$ and $\pi^*$ modes saturate to $0.88$ and $0.95G_0$ respectively, both close to $1G_0$, but 
the $\pi^*$ mode saturates noticeably slower than the $\pi$ mode. 

In order to better understand these saturations, let us take a look
at the conductance in the 'infinite' configuration. This allows us to see a qualitative difference 
between the $\pi$ and the $\pi^*$ modes. The conductance of the modes are splitted into two components:
conductance to the metal and conductance 'through' the nanotube, to the other side, without leaking into the metal. 
The sum of the two conductances, $\sigma_{total}$ is also shown. The quantity $1G_0-\sigma_{total}$ tells us the fraction
of electrons that are reflected for that particular mode.
Taking as an example the (10,10) nanotube, we first consider
a very close metal-nanotube separation corresponding to $d=1\AA$ (fig.3a). The total $\pi$ conductance is
small in this case ($~0.4G_0$), implying that $60\%$ of the current from this mode is reflected (for
$d=0.7\AA$ we obtain $\sim 80\%$ reflection).
Furthermore, all the conductance corresponds to electrons leaking into the metal. On the other hand,
the total conductance for the $\pi^*$ mode is practically $1G_0$, thus $\pi^*$ electrons are almost
not reflected. However, they do not tunnel into the metal easily, and even for $l=120\AA$ still $70\%$ of the
$\pi^*$ electrons tunnel through without leaking into the metal. This completely agrees with the
results of \cite{Ihm}. This kind of behavior is representative of a 'strong coupling regime'. A weak coupling
regime is shown in fig.3b, where a larger nanotube metal separation ($d=3.5\AA$) has been considered.
Now the total $\pi$ conductance is $1G_0$. As $d$ increases, a bigger fraction of the electrons tunnel into the
metal instead of travelling through. The $\pi^*$ electrons, as before, do not tunnel into the metal so easily and need
a longer $l$ than the $\pi$ electrons to leak out. 
Such a $\pi^*$ saturation length is of hundreds of $\AA$ for the (10,10) nanotube
case, but is presumably longer than thousands of $\AA$ for the (40,40) and larger diameter nanotubes. From the
infinite configuration results one concludes that, in a strong coupling regime, the $\pi$ mode gets completely reflected,
but the $\pi^*$ mode still tunnels through with very little leakage into the metal; in a weak coupling regime, none of the
modes is reflected, but they have very different leakage characteristic lengths, the $\pi^*$ mode being slower.
In the case of large diameter nanotubes, this characteristic length is long enough not to be observable experimentally.
The reason for the reflection of the $\pi$ mode in the strong coupling regime is related to the phenomenon of
"maximum saturation conductance" studied in a different context in ref. \cite{Rodero}. This phenomenon states that,
for a finite size contact, the conductance increases as we increase the contact strength from zero, up
to a certain value, after which the conductance decreases if the contact strength is further increased.
In our case this can be understood as follows: in the strong coupling regime, chemical interaction is
taking place between the nanotube and the metal, such that the properties of the dipped carbon atoms
are very different from those of the non-dipped atoms. Therefore, electrons trying to enter from the
non-dipped nanotube part into this new system are mostly reflected. The inability of the $\pi^*$ mode to 
leak into the metal makes it more insensible to strong coupling than the $\pi$ mode. We stress the 
fact that bigger radius nanotubes have larger saturation lengths for the $\pi^*$ electrons, in all coupling regimes. 
By extrapolation of the results up to $200\AA$, we conclude that the saturation length of the $\pi^*$ mode
at a (40,40) nanotube is, in all regimes, longer than thousands of $\AA$. The fact that the conductance curves experimentally
reported in ref.\cite{Frank} saturate in lengths shorter than $100nm$ leads us to think that the 
mode observed in the experiments would be the $\pi$ mode, in a weak coupling regime ($d>1.2\AA$).

In the case of undoped nanotubes the $\pi$ and $\pi^*$ modes are degenerated
at $k_z=k^{nanot}_F$. 
We did not systematically study the effect of doping the nanotubes on 
the conductance. 
Nevertheless, we verified that changing the Fermi level of the nanotube 
up or down makes the saturation length of the $\pi$ mode shorter or longer respectively, and has the 
opposite effect on the saturation length of the $\pi^*$ mode. A shift of $0.2eV$ changed the $\pi$ mode saturation
length by about a factor of 2. However these shifts do not change any of the qualitative behaviors obtained,
and the $\pi$ mode saturates considerably faster than the $\pi^*$ mode in all cases.

Aiming at a better understanding of the tunneling process into the metal, we have studied the contribution
of different electronic modes of the metal to the conductance. In cylindrical coordinates the wavefunctions 
satisfying the boundary condition $\psi_{r\rightarrow r_0+d^+} \rightarrow 0$ for an infinitely 
extended metal in the $z$ direction, are proportional to
\begin{eqnarray}
\psi^{s(c)}_{n,E,k_z}(z,\theta,r) \propto e^{ik_z z} \Theta^{s(c)}_n (\theta) R_{n,E,k_z}(r)
\end{eqnarray}
where
\begin{eqnarray}
\Theta^s_n=\sin (n\theta), \;\;\;   \Theta^c_n=\cos (n\theta),
\end{eqnarray}
$k_z$ is the wavevector in the axial direction (i.e. parallel to the tube), $r_0$ is the
nanotube's redius, and $R_{n,E,k_z}(r)$ is a 
function involving Bessel functions of the first and second kind. It is easy to show that the
nanotube's $\pi$ mode electrons can only couple to metal wavefunctions with two different kinds of
angular dependence, $\Theta(\theta)$: either $\Theta^c_n=0$ or $\Theta^c_n=n_{chir}$, for an 
$(n_{chir},n_{chir})$ nanotube. On the other hand, the $\pi^*$ mode electrons can only couple to metal wavefunctions
of one type of angular dependence: $\Theta^s_n=n_{chir}$. We checked that this is
true by calculating the conductance to the metal when only wavefunctions of one particular angular 
dependence are kept, obtaining a null conductance in all but the three cases mentioned. If we now look at the
available density of states very close to the edge of the metal at the Fermi level, 
for each particular angular dependence type,
we find that it is much bigger in the case of $\Theta_{n=0}$ than for $\Theta_{n=n_{chir}}$ (see fig.4). In fact, in the
limiting case of $n_{chir}\rightarrow \infty$, the d.o.s. for $\Theta_{n=n_{chir}}$ is zero for energies
smaller than $9.4eV$, corresponding to a free electron band shifted by $\hbar^2 k_c^2/2m$ with a wavevector component of
$k_c~0.73a_0$ in the circumpherencial direction. The fact that the nanotube radius is finite makes it possible for 
electrons with $n=n_{chir}$ to still be able to reach the nanotube. The larger the radius, the smaller the available
d.o.s. of the $n=n_{chir}$ electrons at the Fermi level, in relation to the $n=0$ electrons' d.o.s. This might induce one
to think that electrons from the nanotube would mainly flow to $n=0$ electrons in the metal. However, this is not the
case. When we artificially remove all metal states with $n\neq 0$, the resulting conductance is smaller than
$0.3G_0$ in all cases, for any value of $d$. There is a geometrical reason for this, based on the fact that
the coupling between the metal $n=0$ states at the Fermi level and the nanotube $\pi$ states almost cancels out when
the dipped length is larger than three nanotube unit cells (still a fraction of the current 
is able to leak into the metal, despite this cancellation,
due to the metal surface which breaks the translational symmetry). This cancellation takes place when summing terms in
the axial direction, $z$, and is caused by the different periods of oscillation of the metal and nanotube states in the
axial direction. Such a cancellation does not occur for tunneling into the $n=n_{chir}$ states of the metal. We 
explicitly checked that indeed the conductance is carried mostly by these states, and the contribution of the $n=0$
states is less than a $10\%$. This is consistent with the rapidly oscillatory character of the 
conducting states at the metal reported in reference \cite{Ihm}. Nevertheless, tuneling to these states comes mostly from the
nanotube's $\pi$ electrons, rather than the $\pi^*$ electrons, for large radius nanotubes. The reason for an easier tunneling
from the $\pi$ than from the $\pi^*$ mode into the metal $n=n_{chir}$ states seems to be the following. Most metal states
with $n=n_{chir}$ have higher energy than the $\pi$ and $\pi^*$ states of the nanotube, at any value of the
$k_z$ axial wavevector. Therefore, nanotube states are pushed down upon interaction with the $n=n_{chir}$ metal states.
The $\pi^*$ band is quite uniformly pushed down at every $k_z$, because the energy of these states increases with
$k_z$, same as the metal states, thus their energy difference is rather constant as a function of $k_z$. However, the
$\pi$ band states decrease in energy with increasing $k_z$. This behaviour is the opposite of that of the metal states 
to which they can couple to. As a result, $\pi$ states with smaller $k_z$ interact more strongly with the metal states than
those with larger $k_z$.
This results in an effective flattening of the band at the submerged part of the nanotube, with the consequent enhancement 
of the density of states, thus enhancing the tunneling rate of these electrons. Obviously, if the interaction is too strong,
the band is totally pushed below the Fermi level, and the $\pi$ electrons are reflected, as suggested in ref. \cite{Ihm}. 
However, as we have seen, the dipping lengths required for the $\pi^*$ mode to reach a conductance of $1G_0$ in 
large radius nanotubes are far too long to agree with experimental evidence.

The reasonably short saturation distance of the $\pi^*$ mode of the (10,10) nanotube in the weak coupling regime
implies that in this case
one would be able to observe a conductance of $2G_0$, composed of a fast saturation to $1G_0$ superimposed to a slower
saturation reaching $2G_0$ in several tens or hundreds of nm. We propose performing an experiment similar to
that carried out in ref. \cite{Frank} but with smaller diameter nanotubes around ~1nm or less. This would 
enable one to determine whether a "weak coupling"
is actually the experimental case for these tubes. Reporting a fast saturation to $1G_0$ plus 
a slower saturation to $2G_0$ would
mean that we are in the "weak coupling" regime, the first saturation corresponding to the $\pi$ mode and the second
one to the $\pi^*$. On the other hand, in case of a strong coupling regime, 
the expected conductance would be at most $~1G_0$. We are
aware that carrying out such an experiment with small diameter nanotubes might encounter many experimental
difficulties, and is not a trivial task.

In summary, we have shown how the conductance varies as a nanotube is dipped into a free electron metal
with a Fermi energy of 7.13eV. We identify two qualitatively different possibilities, solely
depending on the nanotube-metal spacing: 1) weak coupling, in which 
conduction takes place via the $\pi$ mode, saturating to $1G_0$, while the $\pi^*$ mode saturates more slowly
and is not detectable for nanotube diameters around $\sim5nm$ or wider; 2) strong coupling, in which
the $\pi$ mode is almost totally reflected, and the $\pi^*$ mode still does not leak to the metal in lengths
much longer than 10nm. We conclude that the experimental situation most likely belongs to the weak coupling
regime, and thus the reported $1G_0$ conductance is carried by the $\pi$ mode, and not the $\pi^*$ mode. The 
saturation length of the $\pi^*$ conductance gets shorter for nanotubes of smaller diameter. 
This would allow one to observe the consecutive saturation of the $\pi$ and $\pi^*$ modes in a range of 
tens of nanometers, if an experiment were performed using $\sim 1nm$ diameter nanotubes.

We are indebted to M.P. Anantram from NASA-Ames for fruitful discussions.
We acknowledge Prof. J.P.Lu for his support of the joint
research program between NASA Ames and University of North Carolina at
Chapel Hill.
\end{text}

\begin{figure} 
\mbox{\psfig{file=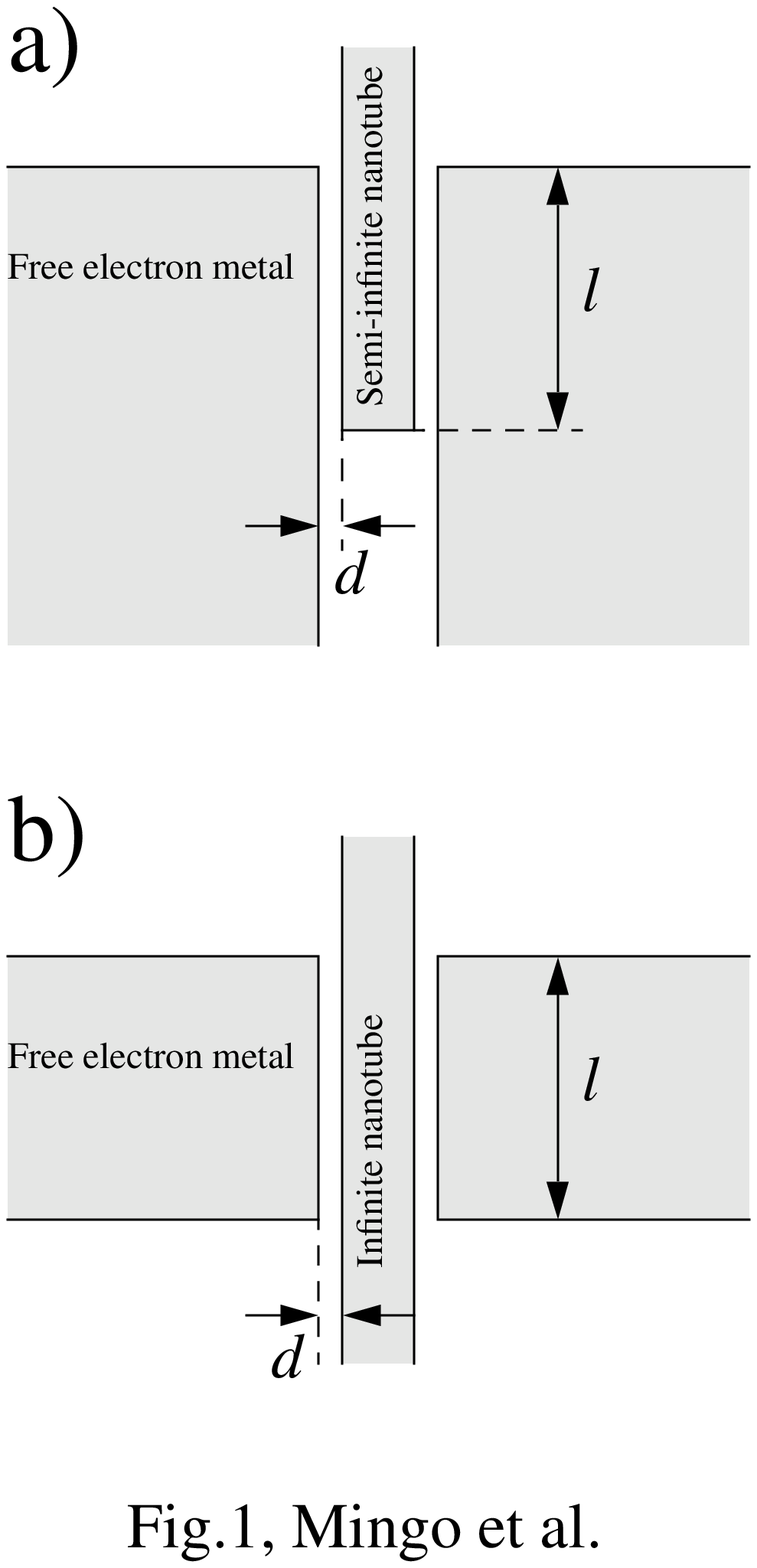,bbllx=100pt,bblly=140pt,bburx=512pt,bbury=734pt,width=8.25cm,clip=}}
\caption{Scheme of the two configurations used. a) Semi-infinite nanotube configuration; b) Infinite
nanotube configuration.}

\label{fig1} 
\end{figure} 

\begin{figure} 
\mbox{\psfig{figure=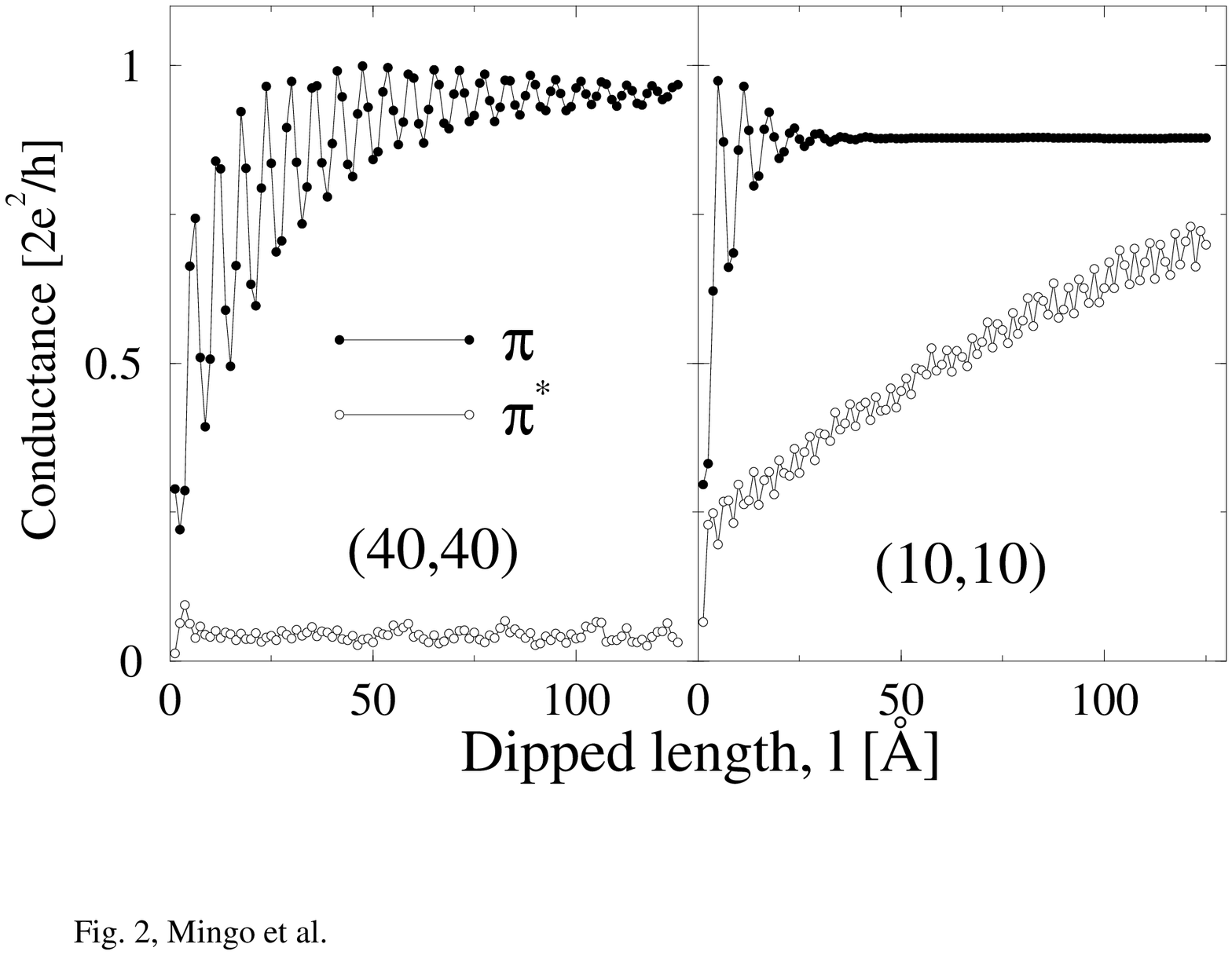,bbllx=35pt,bblly=90pt,bburx=576pt,bbury=457pt,width=8.25cm,clip=}}
\caption{Total conductance from the $\pi$ and $\pi^*$ modes to the metal, as a function
of dipped length, in the semi-infinite configuration. Left: (40,40) nanotube. Right: (10,10) nanotube.
In the two cases $d=1.5\AA$, corresponding to weak coupling regime.}
\label{fig2} 
\end{figure} 

\begin{figure} 
\mbox{\psfig{figure=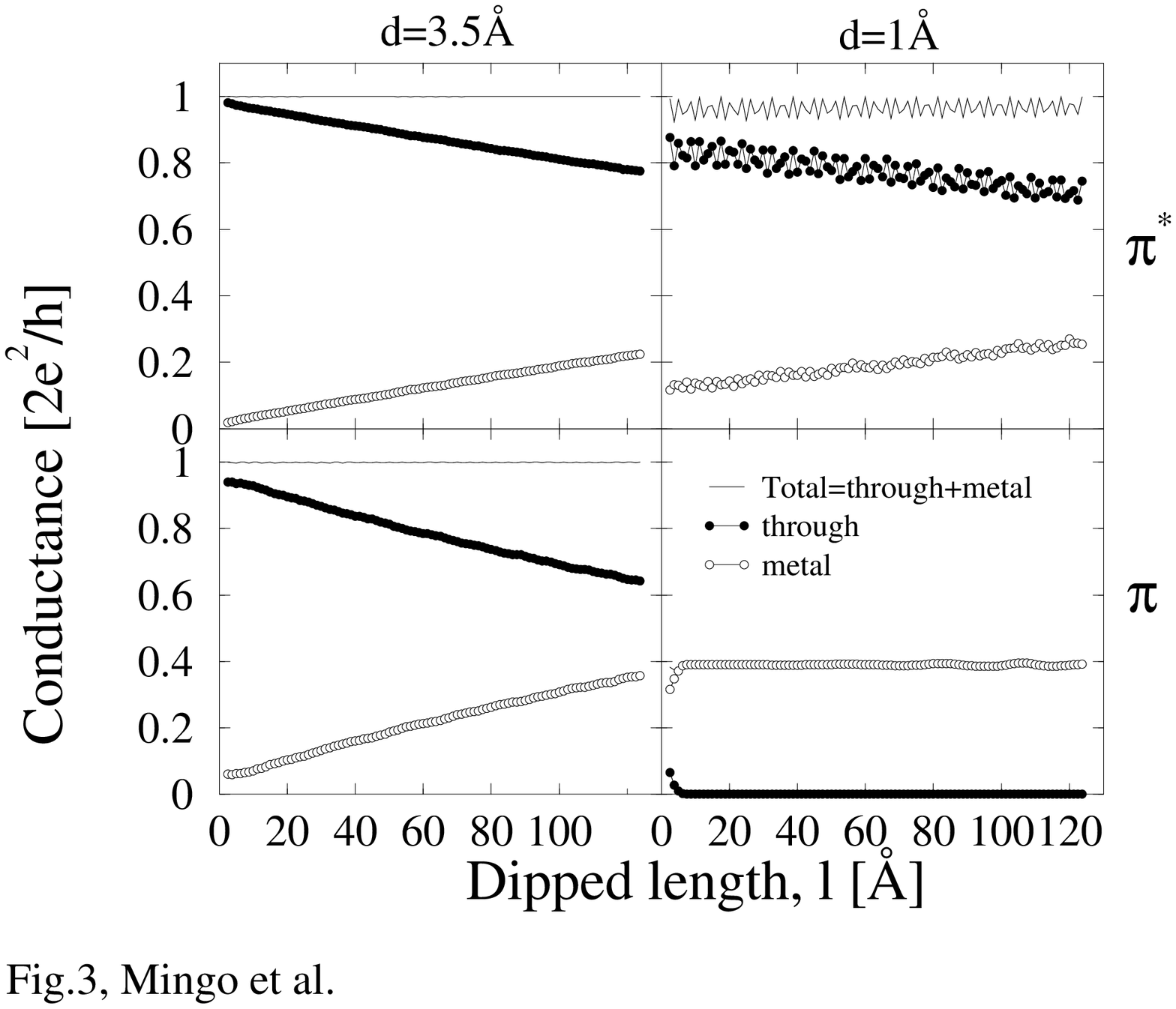,bbllx=35pt,bblly=50pt,bburx=571pt,bbury=482pt,width=8.25cm,clip=}}
\caption{Conductance components of the $\pi$ and $\pi^*$ modes for the (10,10) nanotube in the infinite configuration.
Full circles: conductance 'to metal'; empty circles: conductance 'through' to the other electrode; 
solid line: 'total' = 'to metal' + 'through'. 
Left: weak coupling regime, $d=3.5\AA$; Right: strong coupling regime, $d=1\AA$.}
\label{fig3} 
\end{figure} 

\begin{figure}
\mbox{\psfig{figure=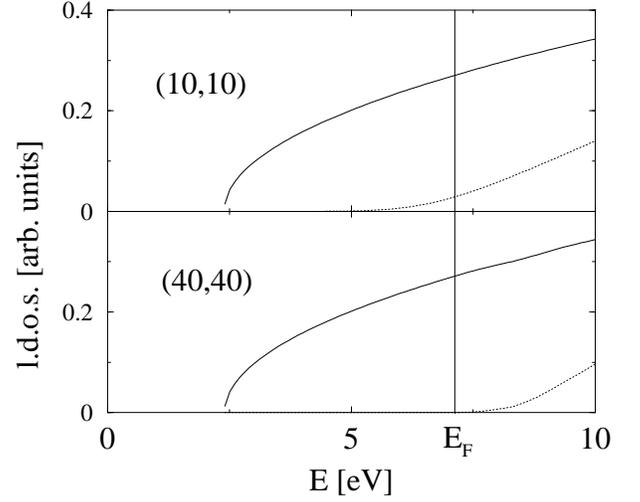,bbllx=40pt,bblly=32pt,bburx=555pt,bbury=460pt,width=8.25cm,clip=}}
\caption{Local densities of states $\rho_{k_z=k_F,r\rightarrow r^+_c}(E)$
at a free electron metal,
close to its cylindrical surface, for the two particular angular dependencies
which are involved in tunneling to the nanotube: $n=0$ (solid line)
and $n=n_{chir}$ (dotted line). Up: case corresponding to a (10,10) nanotube.
Down: for (40,40) nanotube. The density of the rapidly oscillating states
available for tunneling is much smaller in the second case than in the first 
one. This results in a shorter saturation length of the $\pi^*$ mode for
smaller diameters, which might allow its observation for $\sim 1nm$ diameter nanotubes
while preventing it in $\sim 5nm$ or larger nanotubes.
}
\label{fig4}
\end{figure}

\end{document}